\documentclass[showpacs,twocolumn]{revtex4}
 
\usepackage{graphicx,amsmath,subfigure}
\usepackage{amssymb}

\newcommand{\sign}{{\rm sign}}
\newcommand{\avg}[1]{\left\langle{#1}\right\rangle}
\newcommand{\ovl}[1]{\overline{#1}}

\newcommand{\be}{\begin{equation}}

\newcommand{\ee}{\end{equation}}
\newcommand{\bea}{\begin{eqnarray}}
\newcommand{\eea}{\end{eqnarray}}

\begin{document}
 
\title{Generalized minority games with adaptive trend-followers and
contrarians}
 
\author{A. De Martino$^1$, I. Giardina$^1$, M. Marsili$^2$ and
A. Tedeschi$^1$}
 
\affiliation{$^1$INFM-SMC and Dipartimento di Fisica, Universit\`a di
Roma ``La Sapienza'', P.le A. Moro 2, 00185 Roma (Italy)\\$^2$The
Abdus Salam ICTP, Strada Costiera 11, 34014 Trieste (Italy)}
 
\begin{abstract}
We introduce a simple extension of the minority game in which the
market rewards contrarian (resp. trend-following) strategies when it
is far from (resp. close to) efficiency. The model displays a smooth
crossover from a regime where contrarians dominate to one where
trend-followers dominate. In the intermediate phase, the stationary
state is characterized by non-Gaussian features as well as by the
formation of sustained trends and bubbles.
\end{abstract}

 
\maketitle 

Financial markets are known to generate non-trivial fluctuation
phenomena \cite{MS,BP} that are qualitatively reproduced by several
models where agents with prescribed trading rules interact through a
complex mechanism of price formation
\cite{SantaFe,Flo,JDF,LM99,GB03}. Originally devised to get a more
fundamental grasp on the critical behavior of systems of heterogeneous
agents, the minority game \cite{CZ} was able to capture some of the
complex macroscopic phenomenology of markets starting from primitive
microscopic ingredients \cite{CMZ,Hui,CM03}, clarifying the roles of
different factors contributing to the complexity of market
dynamics. Still, many important issues escape a more basic
investigation.

One of these is the interaction of different types of agents. Broadly
speaking, traders can be divided in two groups, namely {\em
contrarians} (or {\em fundamentalists}) and {\em trend-followers} (or
{\em chartists}). The former believe that the market is close to a
stationary state and buy (sell) when they repute the stock to be
underpriced (overpriced), thus inducing anti-correlation in market
returns and holding the price close to its `fundamental' value. The
latter, instead, extrapolate trends from recent price increments and
buy or sell assuming that the next increment will occur in the
direction of the trend, thus creating positive return correlations and
large price drifts (`bubbles').  Chartist behavior, which can also be
driven by imitation, is known to cause market instability
\cite{LM99,M01}. Fundamentalists act instead as a restoring force that
dumps market inefficiencies and excess volatility. It has been argued
\cite{M01,MHet} that contrarians (trend followers) are described by
minority (majority) game players (but see also \cite{AS03,FM03}), and
the analysis of mixed majority-minority games has shown that the
presence of trend followers can severely alter the market's efficiency
\cite{DGM03}. However, agents in these models are committed to either
one of the types, and switching from one group to the other, a key
feature in other models \cite{LM99}, is not allowed.

Here we introduce a class of market games that bypasses this
limitation. We assume that trend-following behavior dominates when
price movements are small, as agents try to anticipate trends, whereas
traders turn to a contrarian conduct when the market becomes
chaotic. This mechanism causes a `feedback' in the dynamics of the
excess demand: when it is small, trend-followers dominate and drive it
to larger values; but once it has become sufficiently large,
contrarians take over and drive it back to smaller values by inducing
anti-correlations. In this way, it is the market that determines
whether trend-following or contrarian strategies gain and there is no
need to employ different payoff functions. We will use the cutoff
between the majority- (where trend-followers win) and the
minority-regime (where contrarians win) as a control parameter to
discriminate fundamentalists- and chartists-dominated phases. The most
remarkable phenomenology occurs, not surprisingly, in-between the two.

Hereafter, we shall use the prefixes $<$- and $>$- for `minority' and
`majority', respectively.

Our basic setup is as follows. At each time step $t$, $N$ agents
receive an information $\mu(t)$ chosen at random from $\{1,\ldots,P\}$
with uniform probability.  Based on $\mu(t)$, agents have to formulate
a binary bid $b_i(t)$ (`buy/sell'). To this aim, each of them is
endowed with $S$ strategies $\boldsymbol{a}_{ig}=\{a_{ig}^\mu\}$
($g=1,\ldots ,S$) that map informations $\mu\in\{1,\ldots,P\}$ into
actions $a_{ig}^\mu\in\{-1,1\}$. Each component $a_{ig}^\mu$ of every
strategy is selected randomly and independently from $\{-1,1\}$ with
equal probability for every $i$, $g$ and $\mu$ at the beginning of the
game and fixed, so that strategies play the role of a quenched
disorder. Finally, each strategy of every agent is given an initial
valuation $p_{ig}(0)$ that is updated at the end of every round.

At each time step, every agent picks the strategy
$\widetilde{g}_i(t)={\rm argmax}_g~ p_{i,g}(t)$ with the largest
valuation and formulates the bid
$b_i(t)=a_{i\widetilde{g}_i(t)}^{\mu(t)}$. The (normalized) excess
demand at time $t$, namely the mismatch between the total demand and
the total supply, is defined as $A(t)=N^{-1/2}\sum_{i=1}^N b_i(t)$,
and strategy valuations are updated according to
\begin{equation}\label{dyn}
p_{ig}(t+1)=p_{ig}(t)+a_{ig}^{\mu(t)} F[A(t)]
\end{equation}
where $F$ is a function embodying the rules with which payoffs are
assigned or the way agent $i$ assesses the performance of his/her
$g^{\rm th}$ strategy (in which case one could also assume that $F$
depends on $i$). In the $<$-game, $F(A)=-A$, so at each time step
strategies suggesting the minority action are rewarded. In the
$>$-game, instead, $F(A)=A$. Here we set
\begin{equation}
F(A)=A-\epsilon A^3\label{M2}
\end{equation}
with $\epsilon\geq 0$. For $\epsilon=0$ one has a pure $>$-Game. Upon
increasing $\epsilon$, the non-linear, $<$-term gains importance, and
for $\epsilon\to\infty$ one obtains a $<$-game with $F(A)\propto
-A^3$.

We want to characterize the steady state of (\ref{dyn}) in the limit
$N\to\infty$, as a function of the relative number of information
patterns $\alpha=P/N<\infty$. In our experiments, we set $p_{ig}(0)=0$
for all $i$ and $g$, and $S=2$, and focused for a start on the
observables $\sigma^2=\avg{A^2}$ and $H=\ovl{\avg{A|\mu}^2}$, where
$\avg{\cdots}$ and $\avg{\cdots|\mu}$ denote time averages in the
stationary state, the latter conditioned on the occurrence of the
information pattern $\mu$, and the over-line stands for an average
over $\mu$'s ($\ovl{\cdots}=(1/P)\sum_{\mu=1}^P \cdots$). Averages
over the distribution of the quenched disorder (the strategies) are
also performed.  As discussed at length in the early $<$-game
literature \cite{MCZ}, $\sigma^2$ measures the magnitude of market
fluctuations (the `volatility'), while $H$ quantifies the
`predictability' of the game, i.e. the presence of exploitable
information: when $H=0$, the winning action cannot be predicted on the
basis of $\mu$. Notice that $\sigma^2=1$ when agents buy and sell at
random.

\begin{figure}[t]
\includegraphics*[width=8.75cm]{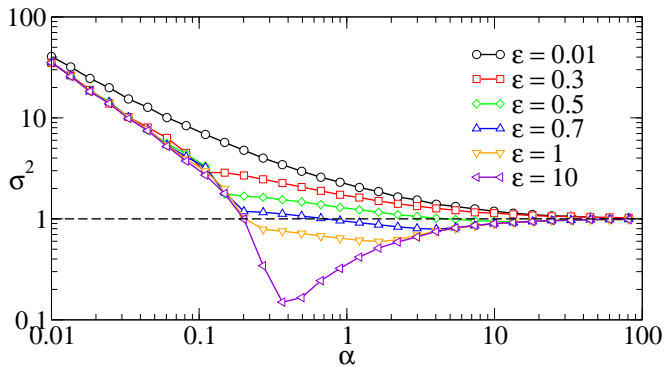}
\caption{\label{M2_s2}Volatility $\sigma^2$ as a function of
$\alpha=P/N$ for different values of $\epsilon$. Simulations performed
with $\alpha N^2=16000$, with averages over 100 disorder samples per
point. A given sample corresponds to a particular realization of the
strategies $\boldsymbol{a}_{ig}$.}
\end{figure}

\begin{figure}[t]
\includegraphics*[width=8.75cm]{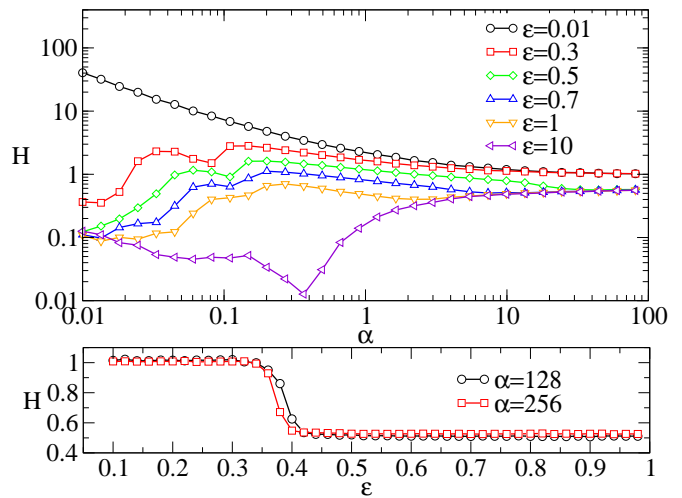}
\caption{\label{M2_H}Predictability $H$ as a function of $\alpha=P/N$
for different values of $\epsilon$ (top) and the reverse
(bottom). Simulation parameters are as in Fig.~\ref{M2_s2}.}
\end{figure}

Results for these quantities are reported in Fig. \ref{M2_s2} and
\ref{M2_H}.
For small $\epsilon$, one recovers as expected a pure $>$-game, with
$\sigma^2=H>1$ for all $\alpha$. As $\epsilon$ increases, the
volatility displays a smooth change to a $<$-regime, with the onset of
a cooperative phase where $\sigma^2$ is better-than-random. When
$\epsilon\to\infty$, a minimum is formed close to the phase transition
$\alpha_c\simeq 0.34$ of the standard $<$-game \cite{CMZ}. The
predictability $H$ shows a more articulated behavior. As $\epsilon$
increases, $H$ becomes smaller than $1$ at low $\alpha$ (as in a
$<$-game), but it still tends to $1$ for large $\alpha$ (as in a
$>$-game). Unfortunately, the low-$\alpha$ behavior is hard to
characterize numerically as a function of $\epsilon$ since reliable
experiments at $\alpha\leq 0.01$ require unrealistic CPU times. For
high-$\alpha$, one can instead identify a sharp transition:
\begin{equation}
\lim_{\alpha\to\infty}H=\begin{cases} 1&\text{for
$\epsilon<\epsilon_c\simeq 0.37$}\\ 1/2&\text{for
$\epsilon>\epsilon_c$}
\end{cases}
\end{equation}
A further increase of $\epsilon$ causes a reduction of exploitable
information. However, no unpredictable regime with $H=0$ is detected
at low $\alpha$ when $\epsilon\to\infty$, at odds with the standard
$<$-game.

\begin{figure}[b]
\includegraphics*[width=8.75cm]{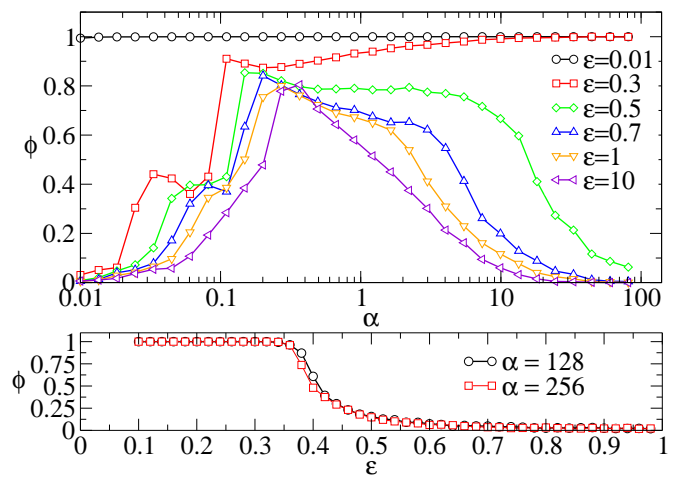}
\caption{\label{M2_phi}Fraction of frozen agents $\phi$ as a function
of $\alpha=P/N$ for different values of $\epsilon$ (top) and the
reverse (bottom). Simulation parameters are as in Fig.~\ref{M2_s2}.}
\end{figure}

Another significant macroscopic observable is the fraction $\phi$ of
``frozen'' agents, that is, of players for which the difference
$p_{i1}(t)-p_{i2}(t)$ between the strategy valuations diverges in the
limit $t\to\infty$, so that they end up using only one of their
strategies. The behavior of $\phi$ is shown in Fig. \ref{M2_phi}.
Again, for large $\alpha$ a sharp threshold separating a $>$-like
regime with all agents frozen ($\epsilon<\epsilon_c$) from a $<$-like
regime where $\phi=0$ ($\epsilon>\epsilon_c$) is found. For large
$\epsilon$, $\phi$ has $<$-game's characteristic shark-fin
shape. Notice that as $\epsilon$ increases $\phi$ decreases, signaling
that it becomes harder and harder for traders to identify an optimal
strategy when the market is dominated by speculators. In the
low-$\alpha$, large-$\epsilon$ phase, our agents are significantly
more likely to be frozen than in a pure $<$-game. This feature,
together with the absence of an unpredictable phase in the same
conditions, is due to the non-standard nature of the $<$-regime in our
model \cite{unpu}.

\begin{figure}[t]
\includegraphics*[width=8.75cm]{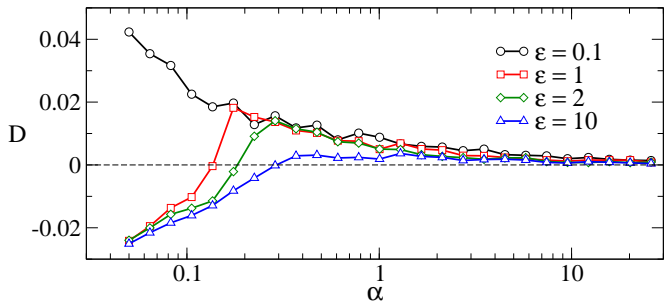}
\caption{\label{D}Normalized correlation function $D$ as a function of
$\alpha=P/N$ for different values of $\epsilon$. Simulation parameters
are as in Fig. \ref{M2_s2}.}
\end{figure}

The fact that $<$-like and $>$-like features can coexist at
intermediate $\epsilon$ can be seen clearly by studying the
correlation \cite{M01,FM03}
\begin{equation}
D=\frac{\avg{A(t)A(t+1)}}{\avg{A(t)^2}}=\frac{\avg{A(t)A(t+1)}}{\sigma^2}
\end{equation}
(see Fig. \ref{D}).
For small $\epsilon$, $D$ is positive, signaling that the market
dynamics is completely dominated by positive correlations (i.e. by
trend-followers). As $\epsilon$ increases, anti-correlations appear at
low $\alpha$. The contrarian phase becomes larger and larger as
$\epsilon$ grows further and for $\epsilon\gg 1$ the market is
dominated by contrarians.

We can shed some light on the crossover from the $<$- to the
$>$-regime at large $\alpha$ for $S=2$ drawing inspiration from
\cite{CM99}. Let us define, for each agent, the strategy valuation
difference $y_i(t)=\frac{1}{2}[p_{i1}(t)-p_{i2}(t)]$, and note that
the Ising spin $s_i(t)\equiv\sign[y_i(t)]$ determines the strategy
that agent $i$ chooses at time $t$.  It is simple to show that
\begin{equation}
v_i\equiv\avg{y_i(t+1)-y_i(t)}=\ovl{\xi_i^\mu\avg{F(A)|\mu}}
\end{equation}
where ${\xi}_i^\mu=\frac{1}{2}({a}_{i1}^\mu- {a}_{i2}^\mu)$. If
$v_i\neq 0$, then $y_i(t) \sim v_i t$ and $s_i(t)$ tends
asymptotically to $\sign(v_i)$: there is a well defined preference
towards one of the two strategies and the agent becomes frozen. This
is what happens for $\epsilon\ll 1$, i.e. in the $>$-game regime.
Here, $A(t)$ is a function of $\mu(t)$ only (because all agents are
frozen), therefore $\avg{A^2|\mu}=\avg{A|\mu}^2$ and $\sigma^2=H$.
For large $\alpha$, when the agents' strategic choices are roughly
uncorrelated \cite{nota2}, we can approximate $A(t)$ with a Gaussian
rv with variance $H$. By virtue of Wick's theorem, this implies that
$\avg{A^3|\mu}\simeq 3H\avg{A|\mu}$, so
\begin{equation}
v_i\simeq (1-3\epsilon H)\ovl{\xi_i^\mu\avg{A|\mu}}
\end{equation}
If $1-3\epsilon H>0$, the agents' spins will freeze on the $>$-game
solution $s_i=\sign(\ovl{\xi_i^\mu\avg{A|\mu}})$, which is unstable
for $1-3\epsilon H\leq 0$. Given that $H=1$ for large $\alpha$, we see
that the crossover from the $>$- to the $<$-regime takes place at
$\epsilon\simeq 1/3$ for $\alpha\gg 1$. This estimate is significantly
close to the numerical value of $\epsilon_c\simeq 0.37$. A similar
argument can be run from the $<$-game side, where, at large $\alpha$,
$A(t)$ can be approximated with a Gaussian random variable with
variance $\sigma^2$ (in this case different from $H$), so that
$\avg{A^3|\mu}\simeq 3\sigma^2\avg{A|\mu}$. Arguing as before, one
finds that the stability condition of a $<$-game like solution is
$1-3\epsilon\sigma^2<0$. Given that $\sigma^2=1$ for large $\alpha$,
we find that the solution is $<$-game like for $\epsilon>1/3$ when
$\alpha\gg 1$.

\begin{figure}[t]
\includegraphics*[width=8.75cm]{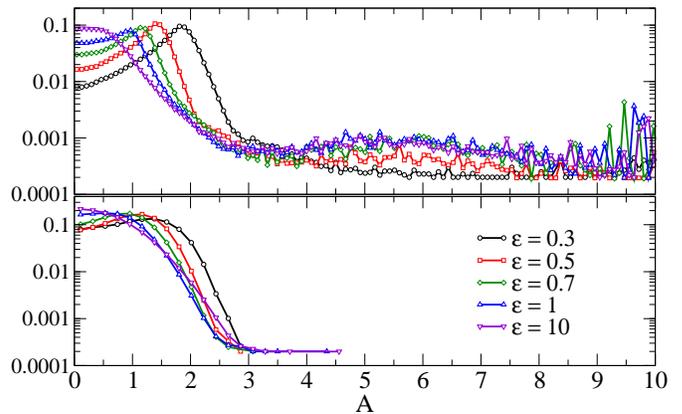}
\caption{\label{Pa005a2}Probability distributions $P(A)$ of $A>0$ for
different values of $\epsilon$ for $\alpha=0.05$ (top) and $\alpha=2$
(bottom). Simulation parameters are as in Fig. \ref{M2_s2}.}
\end{figure}

Unfortunately, the above argument is not valid at small $\alpha$ since
$A(t)$ acquires strong non-Gaussian statistics, so the quality of the
approximation gets worse and worse as $\alpha$ decreases. To see this,
let us inspect the probability distribution $P(A)$ of $A(t)>0$ as a
function of $\epsilon$ in the regimes of large and small $\alpha$ (see
Fig. \ref{Pa005a2}).
For $\alpha=2$ and $A$ not too large, one finds roughly that $\log
P(A)\propto A^2-b A^4$ ($b=$constant) with a weak dependence on
$\epsilon$.  For $\alpha=0.05$, instead, $P(A)$ is considerably more
sensitive to the the value of $\epsilon$ and cannot be fitted by a
simple form as before. In this regime, where the contribution of
frozen agents is small, we expect the system to self-organize around a
value of $A$ such that $F(A)=0$: indeed one can see from
Fig. \ref{Pa005a2} that the peak of the distribution moves as
$1/\sqrt{\epsilon}$. Besides, as $\epsilon$ increases, large excess
demands occur with a finite probability. The emergence of such `tails'
in $P(A)$, while not power-law, is a clear non-Gaussian signature.

\begin{figure}[t]
\includegraphics*[width=8.75cm]{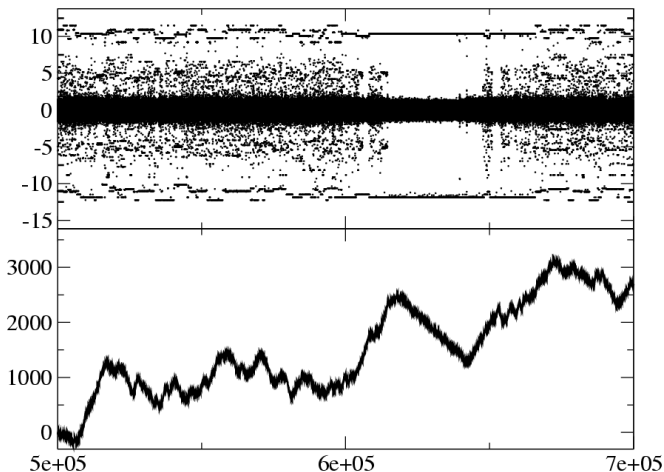}
\caption{\label{SS}Single realization of market dynamics at
$\alpha=0.05$ and $\epsilon=1$. Top: $A(t)$ vs $t$. Bottom: $R(t)$ vs
$t$.}
\end{figure}

In the light of these findings, it is interesting to inspect the
typical market dynamics in the non-Gaussian regime. In Fig. \ref{SS} a
single realization of the game at $\alpha=0.05$ and $\epsilon=1$ is
displayed. In particular, we show the time series of excess returns
and the time series of the price $R(t)=\sum_{t'\leq t} A(t')$ in the
steady state. One can clearly see that while the market is mostly
chaotic and dominated by contrarians, `ordered' periods can arise
where the excess demand is small and trends are formed, signaling that
chartists have taken over the market. A detailed analysis clarifies
that the spikes in $A(t)$ occur in coordination with the transmission
of a particular information pattern $\nu$ to which the market responds
by generating large excess demands. This phenomenon is reminiscent of
the retrieval of stored patterns in neural networks, and was also
found in $>$-games \cite{KM}, although in the present case $A(t)$ is
not of order $\sqrt{N}$. However, the `recalled' pattern $\nu$ changes
with time, and each pattern can be `active' for many time steps in a
row and then quiesce for just as long during a single run. These
features make the dynamics strongly sample-dependent, in a way that is
reminiscent of another recently studied variation of the $<$-game
\cite{CMD}.

In summary, we have introduced a class of minority games in which the
market determines whether, at each time step, contrarians or trend
followers profit, showing that market-like phenomenology emerges when
the competition between the two groups is stronger. This work raises
many further questions, concerning the presence of phase transitions,
the change induced by using real market histories instead of random
information and, in particular, the extension of this model to
grand-canonical settings (the latter appears to be especially
promising as some empirical facts such as volatility clustering can
emerge only in games where the number of traders fluctuates in time
\cite{CM03,GBM,BGM}).  Work along these lines is currently in
progress.

This work was partially supported by the EU EXYSTENCE network and by
the EU Human Potential Programme under contract HPRN-CT-2002-00319,
STIPCO. A.D.M. gratefully acknowledges The Abdus Salam ICTP for
hospitality.


\begin{thebibliography}{99}
\bibitem{MS}R.N. Mantegna and H.E. Stanley, {\it An introduction to
econophysics} (Cambridge University Press, 1999)
\bibitem{BP}J.-Ph. Bouchaud and M. Potters, {\it Theory of financial
risks} (Cambridge University Press, 2000)
\bibitem{SantaFe}W.B. Arthur, J.H. Holland, B. LeBaron, R. Palmer and
P. Tyler, in W.B. Arthur, S.N. Durlauf and D.A. Lane (Eds.), {\it The
economy as an evolving complex system II} (Addison Wesley, Reading,
MA, 1997)
\bibitem{Flo}J.D. Farmer and A.W. Lo, Proc. Nat. Acad. Sci. {\bf 96}
9991 (1999)
\bibitem{JDF}J. D. Farmer, Comp. Sci. Eng. {\bf 1} 26 (1999)
\bibitem{LM99}T. Lux and M. Marchesi, Nature {\bf 397} 498 (1999)
\bibitem{GB03}I. Giardina and J.-Ph. Bouchaud, Eur. Phys. J. B {\bf
31} 421 (2003)
\bibitem{CZ}D. Challet and Y.-C. Zhang, Physica A {\bf 246} 407
  (1997). See also http://www.unifr.ch/econophysics/minority.
\bibitem{CMZ}D. Challet, M. Marsili and Y.-C. Zhang, Physica A {\bf
276} 284 (2001)
\bibitem{Hui}P. Jefferies, M.L. Hart, P.M. Hui and N.F. Johnson,
Eur. Phys. J. B {\bf 20} 493 (2001)
\bibitem{CM03}D. Challet and M. Marsili, Phys. Rev. E {\bf 68} 036132
(2003)
\bibitem{M01}M. Marsili, Physica A {\bf 299} 93 (2001)
\bibitem{MHet}M. Marsili, in A. Kirman and J.P. Zimmermann (Eds.),
{\it Economics with heterogeneous interacting agents} (Springer,
Berlin, 2001)
\bibitem{AS03}J. V. Andersen and D. Sornette, Eur. Phys. J. B {\bf 31}
141 (2003)
\bibitem{FM03}F. Ferreira and M. Marsili, Preprint cond-mat/0311257.
\bibitem{DGM03}A. De Martino, I. Giardina and G. Mosetti, J. Phys. A
{\bf 36} 8935 (2003) 
\bibitem{MCZ}M. Marsili, D. Challet and R. Zecchina, Physica A {\bf
280} 522 (2000)
\bibitem{unpu}If one performs a similar analysis with the piecewise
linear payoff function $F(A)=[\theta(\eta-|A|)-\theta(|A|-\eta)]A$ in
place of (\ref{M2}), one observes that, at low enough $\eta$, $H=0$
for $\alpha<\alpha_c(\eta)$ with $\alpha_c(0)=0.3374\ldots$ and that
$\alpha_c$ decreases as $\eta$ is increased, similarly to what happens
in the mixed majority-minority game \cite{DGM03}. Furthermore,
analyzing the stationary behavior in the limit $\alpha\to\infty$, one
finds that a sharp transition between a $<$-like (low $\eta$) and a
$>$-like (high $\eta$) regime takes place at $\eta_c\simeq 1.41$.
\bibitem{CM99}D. Challet and M. Marsili, Phys. Rev. E {\bf 60} R6271
(1999)
\bibitem{nota2} When the relative number of information patterns
$\alpha=P/N$ is large the $2 N$ randomly selected strategies are
typically very different from each other.
\bibitem{KM}P. Kozlowski and M. Marsili, J. Phys. A {\bf 36} 11725
  (2003)
\bibitem{CMD}D. Challet, M. Marsili and A. De Martino, Preprint
cond-mat/0401628.
\bibitem{GBM}I. Giardina, J.-Ph. Bouchaud and M. M\'ezard, Physica A
{\bf 299} 28 (2001)
\bibitem{BGM}J.-Ph. Bouchaud, I. Giardina and M. M\'ezard,
Quant. Finance {\bf 1} 212 (2001)
\end{thebibliography}
\end{document}